# LINEAR ARRAYS OF NON HOMOGENEOUS Cu SITES IN THE CUO$_2$ PLANE, A NEW SCENARIO FOR PAIRING MECHANISMS IN A CURRUGATED-IRON-LIKE PLANE

Antonio Bianconi

*University of Rome "La Sapienza", Department of Physics, P. A. Moro 2, 00185 Roma, Italy*
*www.superstripes,com*      Email : antonio.bianconi@roma1.infn.it

ABSTRACT

Experimental results obtained by using x-ray absorption spectroscopy show that the configurations of Cu sites in the CuO$_2$ plane of Bi 2:2:1:2 high T$_c$ superconductors are not homogeneous. Different Cu sites are characterized by short 2.3 Å and long 2.45 Å Cu-O(apical) distances. The linear arrays of different Cu sites forming domains with a corrugated-iron like shape is proposed to be a key characteristic of superconducting domains in the CuO$_2$ plane. The wavelength of the modulation is close to the superconducting coherence length. The ordering of the distorted Cu sites is suggested to be evidence for ordering of polarons driven by the pseudo Jahn Teller electron lattice interaction. The Cu L$_3$ XAS experiments on Bi 2:2:1:2 system indicate that for δ =9% electronic states added by doping 4±2 % have the a$_1$ symmetry (i.e. with Cu $\underline{3d}_{3z^2-r^2}$, the combination of O (planar) $2p_{x,y}$ orbital with a$_1$ symmetry $\underline{L}(a_1)$, and O(apical) $2p_z$ orbital character) and 15±2% have the b$_1$ symmetry ($\underline{3d}_{x^2-y^2}$ and the combination of O (planar) $2p_{x,y}$ orbital with b$_1$ symmetry $\underline{L}(b_1)$). This new scenario supports the pairing mechanisms for high T$_c$ superconductivity in the presence of two components: 1) the more delocalized component with b$_1$ symmetry and 2) the more localized component, with partially a$_1$ symmetry associated with different parts of the Fermi surface.

## 1. Introduction

So far the two dimensional CuO$_2$ plane where pairing takes place in high T$_c$ cuprate superconductors[1] has been considered to be homogeneous, i.e. the Cu site structure configuration (the Cu square plane in electron doped systems, the square pyramid in YBaCuO and Bi 2:2:1:2 systems and the elongated bipyramid in doped La$_2$CuO$_4$) is assumed to be the same at all Cu sites, In this work we want to point out that the Cu site structure is found to be not homogeneous in Bi 2:2:1:2 by EXAFS and diffraction works[2] and it is modulated with a period close to the superconducting coherence length. The different Cu sites are correlated both with ordering of polarons in the $CuO_2$ plane and with ordering of dopants: interstitial oxygen, defects, substituted ions. The presence of these ordered domains is difficult to be detected because the size of the ordered domains can be so small 10-30 Å to escape to be detected by electron, x-ray and neutron diffraction experiments and the system could appear to be a solid solution on a large scale. Only where ordered domains with size larger than 100 Å are formed they can be detected by diffraction and only if large samples giving intense superstructure spots are obtained the x-ray and neutron diffraction methods are able to solve the structure of the different Cu sites contributing to the superstructure. In high $T_c$ superconductors the short coherence length 10-30 Å implies that the pairing could take place in small ordered domains and their ordering could be a key requirement for the pairing mechanism

X-ray absorption experiments by using synchrotron radiation probe the Cu site structure without requiring long range order. The extended x-ray absorption fine structure (EXAFS) probes the Cu-O inter-atomic distances with a 0.02 Å accuracy and with a measuring time of $10^{-16}$ sec. Therefore it provides a distribution of interatomic distances due to both static and dynamic contributions. EXAFS investigations[3] have found the presence of two Cu-O(apical) distances, 0.13 Å apart, in $YBa_2Cu_3O_{6.5}$, $YBa_2Cu_3O_7$, and about 0.2 Å apart in $TlBa_2Ca_3Cu_3O_{11}$ while diffraction data have reported a single Cu-O(apical) average distance. Anomalies on the variation of the Cu site structure at $T_c$ were found by EXAFS and XANES.[3, 4-6] The pulsed neutron diffraction experiments[7] probing the local pair distribution function without requirements for long range order have found evidence for a split position of the apical oxygen and changes in the local arrangement of oxygen atoms in the $CuO_2$ plane across $T_c$. These results are not in agreement with x-ray and neutron diffraction data which probe the average atomic coordinates over domains of the order of 100-200 Å size.

In few superconducting systems Bi 2:2:1:2[8], $YBa_2Cu_3O_{6.5}$, $La_2CuO_{4.09}$[9] electron diffraction experiments show the presence of a superstructure due to ordering of dopant ions over domains larger than 100 Å, on the contrary in other systems, like LaSrCuO $YBa_2Cu_3O_7$, no superstructure has been detected indicating that if domains are present their size is not larger than 20-30 Å. However where the incommensurate superstructure is observed by diffraction methods the intensity of the superstructure spots are usually so weak that it is hard to extract the coordinates of the atoms of the different Cu sites contributing to the superstructure. This is the reason why only few diffraction experiments have revealed the different Cu site configurations by solving the superstructure.

We have recently investigated the local Cu site structure in Bi 2:2:1:2 system[2] by EXAFS in order to solve the controversy between EXAFS and diffraction methods. We have selected the Bi 2:2:1:2 system where the superstructure in neutron an x-ray diffraction has been solved by several groups[10-12]. In the Bi 2:2:1:2 system the Cu site structure modulation is stabilized by the mismatch between the BiO rock salt layers and the perovskite layers[10-12]. The crystals show the $\lambda a \times 1b$ incommensurate superstructure where the wavelength of the supestructure $\lambda$ is found to be about 4.75. In superconducting samples where the hole doping has been controlled by Y to Ca substitution[13] and in samples prepared in Ar atmosphere with different oxygen content[14] the period $\lambda$ has been found to change in the range 4.6 - 5 correlated with doping and the critical temperature. On the contrary $\lambda$ is found to be around 4 in the insulating phase.

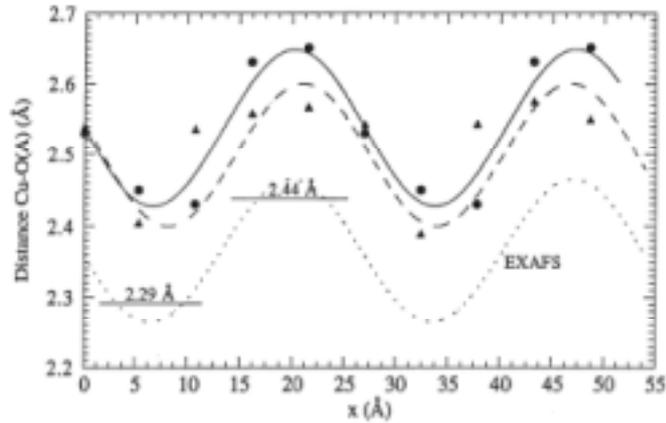

**FIG. 1** The modulation of the Cu-O(apical) bond distances along the (100) direction due to the 5x1 superstructure obtained by x-ray diffraction ref.10 (dots), and to the 4.75x1 superstructure solved by neutron diffraction ref. 11 (triangles) and as obtained by our EXAFS work, ref. 2.

**2. Non homogeneous Cu sites**

*a) The modulated Cu apical oxygen distance.*

Two average distances for the Cu-apical oxygen bond (Cu-O(apical)) have been found by EXAFS analysis: a short bond 2.29 Å with a probability of 60±10% and a long bond 2.44 Å with a probability of 40 ± 10% as shown in Fig. 1. The separation between the two positions is 0.15 ± 0.04 Å which is similar to that found in $YBa_2Cu_3O_7$ [3]. Because each Cu site is coordinated by only one apical oxygen two interpretations are possible: first, the apical oxygen moves in a double well potential, determined by a structural instability[3]; second, the Cu sites are not homogeneous i.e. there are two sets of different Cu sites with long and short distances. We have found that this second interpretation is in agreement with the three diffraction experiments[10-12] that solved the superstructure in Bi 2:2:1:2 crystals.

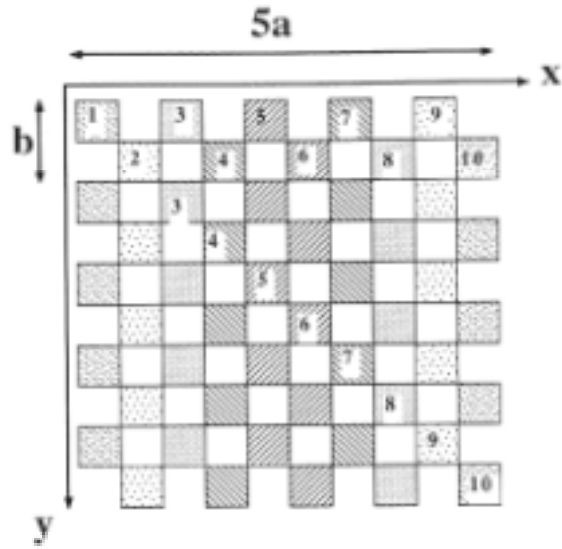

**FIG. 2.** Pictorial view of the $CuO_2$ plane with a 5x1 superstructure where the different $CuO_4$ square planes characterized by short and long Cu-O(apical) bonds are indicated.

In Fig. 1 we report the Cu-O(apical) (or Cu-O(A)) distances for different Cu sites along the (100) direction as found by different authors. The Cu-O(apical) distance has been found to be modulated over 5 or 4.75 crystalline unit cells i.e. with a period of 26-27 Å as it is shown in Fig. 1 where the Cu-O(apical) distances reported by ref. 10 and 11 are plotted and they have been fitted with a formula $d = d_0 + A \cos[2\pi/\lambda(x+\phi)]$. All authors agree on the amplitude of modulation A = 0.1 Å but they give different average values $d_0$ = 2.53 Å at 290K, $\lambda$ = 5a[10], $d_0$ = 2.47 Å at 90K, $\lambda$ = 4.75a [11]. At the present status of EXAFS data analysis it is not possible to solve the five different distances expected for the 5x1 superstructure, therefore we have assumed only two distances in the EXAFS analysis and the effective Debye Waller factors take account of the distribution of the Cu-O distances. Therefore the two distances found by the present EXAFS work indicate the presence of the distribution of Cu-O(apical) distances over a range of 0.15 Å in agreement with diffraction data but around a shorter average Cu-O(apical) bond length of 2.37 Å. The joint interpretation of EXAFS and diffraction data clarifies that the short and long Cu-O(apical) distances are not due to the apex oxygen instability between two positions but it is due to the presence of different Cu site structure configurations in the $CuO_2$ plane. In Fig. 2 we present a pictorial view of the $CuO_2$ plane with a 5x1 superstructure with the linear arrays of different Cu square pyramids. In Fig. 3 a pictorial view of the modulation of the Cu-O(apical) distance in two dimensions is presented. Because the Cu-O(apical) distance modulates the local electronic structure of the Cu site, the $CuO_2$ plane looks more like a corrugated-iron foil, where the different Cu sites are aligned along the linear grooves, than like a flat layer as it was considered until now.

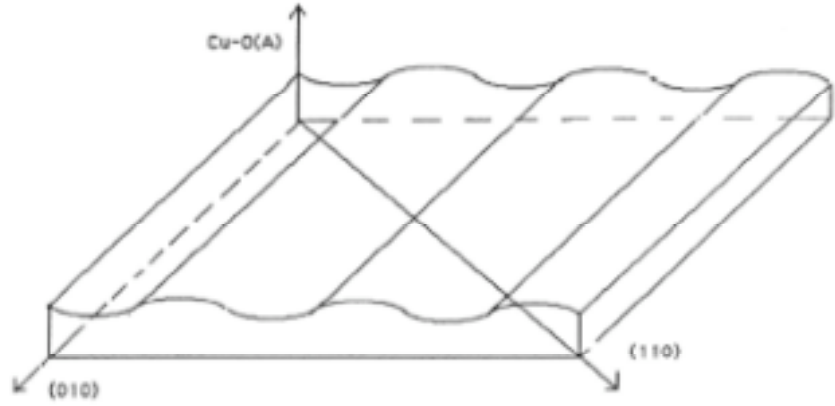

FIG. 3. Pictorial view of the $CuO_2$ plane where the Cu-O(apical) distance is modulated only in one direction, giving a corrugated iron like surface for apical oxygen which implies that also the electronic structure of the $CuO_2$ plane is modulated like a corrugated-iron foil.

In order to investigate if these results are unique for Bi 2:2:1:2 or indicate a common feature for the high $T_c$ superconductors we have analyzed the case of the $YBa_2Cu_3O_{6.5}$. In fact also in this case many authors have reported the presence of a 2x1 superstructure due to linear and empty chains Cu(1)-O in the basal plane in the ortho II phase[15]. We have analyzed the average Cu(2)-O(apical) distance reported by Cava et al.[16]. In Fig.4 the fit of Cu(2)-O(apical) distance with two different distances in the range of oxygen concentrations 6.35-6.75 indicate that it is possible to assign the average Cu(2)-O(apical) distance measured by the diffraction experiments that have not solved the superstructure to the presence of two distances 2.27 Å and 2.45 Å where the probability of the short distance is taken to be given by the relative presence of four coordinated Cu(1) ions in the basal plane[17] increasing with oxygen doping. This result[18] is in agreement with EXAFS data showing the presence of two Cu(2)-O(apical) distances 0.13 Å apart. This result is in fully agreement with the results of Burlet et al.[19] who succeeded to solve the 2x1 superstructure in $YBa_2Cu_3O_{6.5}$ by measuring neutron diffraction data for a very large crystal. The results of Burlet et al. show the presence of sites with long 2.42 Å and short 2.32 Å Cu (apical) distances which reconciles the EXAFS and diffraction data for the presence of two different Cu-O(apical) distances within the experimental errors of 0.04 Å.

It is therefore possible to conclude that for the two superconducting crystals, Bi 2:2:1:2 and $YBa_2Cu_3O_{6.5}$, where the superstructure was solved by neutron diffraction and the local structure was investigated by EXAFS the two methods converge showing the presence of linear arrays of Cu sites with short and long Cu-O(apical) distances. In conclusion the linear arrays of different Cu sites configurations make the symmetry of the $CuO_2$ plane more like that of a corrugated iron foil as it is shown n Fig.3 than like a flat homogeneous two dimensional layer.

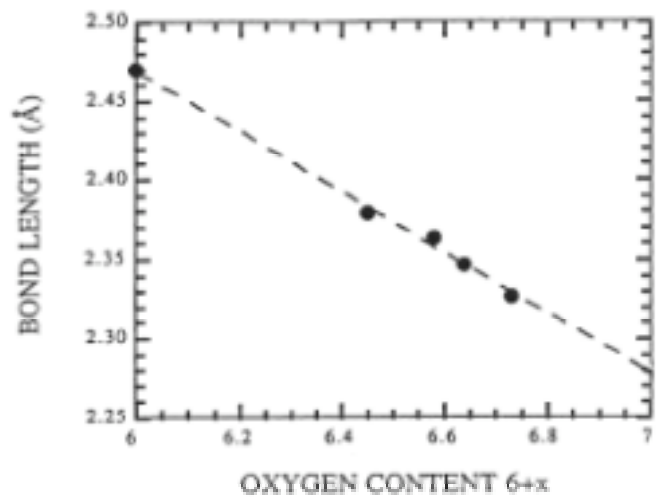

**FIG. 4** The average Cu(2)-O(apical) bond distances measured by Cava et al.[16] in $YBa_2Cu_3O_{6+x}$ fitted with the combination of two Cu(2)-O(apical) distances 2.27 Å and 2.4 Å where the relative weight of the short 2.27 Å distance is taken to be proportional to the number of fourfold coordinated Cu1 ions in the basal plane[17].

*b) The distortion of the CuO$_4$ square plane*

Here we want to point out that the structural investigations indicate that the distortion of the CuO$_4$ square plane is a key requirement for superconductivity. From the EXAFS investigation of Bi 2:2:1:2 we have found two short 1.88 Å and two long 1.95 Å Cu-O(planar) distances. This result is in qualitative agreement with all diffraction data[10-12] and in quantitative agreement with neutron diffraction data at 90K and 55K on a crystal with a =5.397 Å, b =5.401 Å and c =30.716 Å where the average Cu-O(planar) distances have been found to be 1.88 and 1.94 Å. Therefore this result confirms the distortion of the CuO$_4$ square planes in Bi 2:2:1:2 also if the **a** and **b** axis are very close.

The present results on the distortion of the Cu square plane on Bi 2:2:1:2 makes this system similar to other superconducting systems. In fact it seems that many crystallographic investigations of superconductors indicate that the orthorhombic distortion is a key requirement for high T$_c$ superconductivity. The tetragonal to orthorhombic transition has been found at the insulating to metal transition in YBa$_2$Cu$_3$O$_{6+x}$, around the oxygen concentration of 6.3[17]. At low temperature in the metallic superconducting phase La$_{1.85}$Sr$_{0.16}$CuO$_4$ shows the orthorhombic distortion. On the other hand the orthorhombic to tetragonal transition in La$_{2-x}$Sr$_x$CuO$_4$.[20] at low temperature above x=0.21 Sr doping has been found to be correlated with the transition from the superconducting phase to the normal metal phase. Moreover it is well established that the crystallographic transition from the low temperature orthorhombic phase to the tetragonal phase at x=0.12 in La$_{2-x}$Ba$_x$CuO$_4$ suppresses the superconductivity.

*c) The Cu displacement from the oxygen plane.*

A characteristic feature of the superconducting phase of YBa$_2$Cu$_3$O$_{6+x}$ for x > 0.3 is the buckling (or dimpling) angle β ~ 8° degrees formed by O(planar)-Cu-O(planar) due to the displacement of the Cu ion from the coordinated oxygen O(planar) plane. By comparison of the Cu-O(planar) distances found by EXAFS and the value of the a axis we can deduce an average buckling (or dimpling) angle of < β > = 4° degrees in Bi 2:2:1:2. This result is in agreement with the diffraction data showing that the buckling (or dimpling) angle is modulated in different Cu sites between 0° and 8° and the displacement of the Cu ion from the plane of the four oxygen ions can be as large as h ~ 0.25 Å. The Cu displacement is modulated with the superstructure: it is close to zero at the Cu sites with long Cu-O(apical) distance and it is larger h ~ 0.25 Å for the Cu sites with the shortest Cu-O(apical) distance.

*d) The electron - lattice interaction and the pseudo Jahn Teller coupling.*

The crystalline structures of parent Cu(II) insulating compounds of the high $T_c$ superconductors exhibit the Cu site structure configurations typical of the Cu(II) Jahn-Teller ions with elongated $CuO_6$ octahedra, as in $La_2CuO_4$, square pyramids, as in $Bi_2YSr_2Cu_2O_{8+\delta}$ and $YBa_2Cu_3O_6$, and square planes, as in $Nd_2CuO_4$. The Jahn Teller effect characteristic of the Cu(II) ions, removes the degeneracy of the upper $E_g$ states $3d_{x^2-y^2}$, ($m_\ell=2$) and $3d_{3z^2-r^2}$, ($m_l=0$) in the octahedral $O_h$ symmetry by pushing up the energy of the $3d_{3z^2-r^2}$, $m_\ell=0$, and lowering the energy of the $3d_{x^2-y^2}$, $m_\ell=2$, by reducing the $O_h$ symmetry with an elongation of the Cu-O bond in the **z** direction, or in the extreme cases by pushing away one or two oxygen ions forming Cu sites with a square pyramid or a square plane coordination. Therefore in the divalent cuprate perovskite the single hole per unit cell is stabilized in the in the Cu 3d derived states with the component of orbital momentum $3d_{x^2-y^2}$, $m_\ell=2$.

The distortions of the $CuO_4$ square planes in doped superconductors can be interpreted as a pseudo Jahn-Teller type distortions[21, 22] mixing the $m_\ell=2$ with $m_\ell=0,1$ states, therefore the transition from the tetragonal insulating phase to the orthorhombic superconducting phase can be correlated with the mixing of the $3d_{x^2-y^2}$ with $3d_{3z^2-r^2}$ (and $3d_{xz}$, $3d_{yz}$) hole states. In Fig. 5 the distortions of the Cu square plane mixing the $3d_{x^2-y^2}$ with $3d_{3z^2-r^2}$ hole states are shown. Therefore the distortions of the Cu square plane induced by doping can be classified as due to the increase of the pseudo JT distortions associated with the $3d^9\underline{L}$ states formed by doping.

The variation of the Cu-O(apical) distance induces the variation of the energy splitting $\Delta_{JT}$ between the $3d_{x^2-y^2}$ and $3d_{3z^2-r^2}$ states (called also the Jahn Teller splitting) and therefore by decreasing the Cu-O(apical) distance the mixing between the $3d_{x^2-y^2}$ with $3d_{3z^2-r^2}$ hole states in the $CuO_2$ planes increases.

The dimpling angle β enters in the expression for the coupling of the electronic states with the local lattice deformations in the theory of the pseudo-Jahn Teller effect[21]. In fact the electron lattice interaction of the pseudo JT type is given by $\lambda = g(Q) f(\Delta_{JT}) h(\beta)$ where the configuration

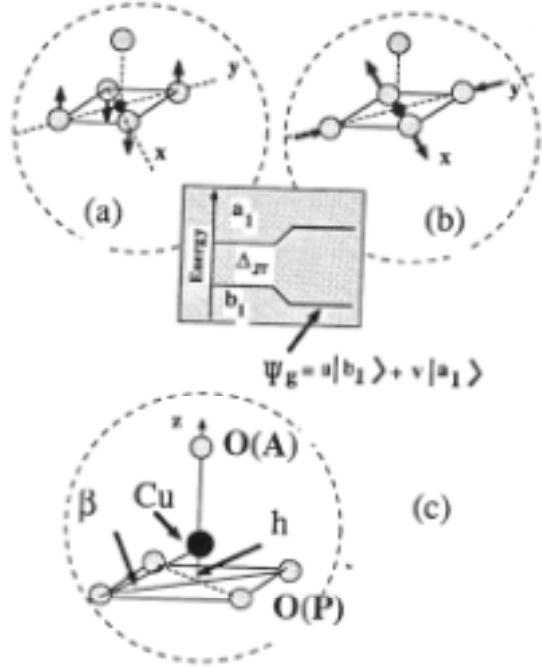

**FIG. 5** The square plane distortions mixing the $a_1$ and $b_1$ states separated by the energy $\Delta_{JT}$ according to the pseudo Jahn Teller (JT) effect: a) The out of phase vibrations of in plane oxygen O(P) and b) the rhombic distortion of the square plane. The energy $\Delta_{JT}$ is a function of the Cu-O(apical) distance and the electron lattice coupling constant is a function of the displacement h of the Cu ion from the oxygen plane or the dimpling angle $\beta$, panel c.

parameter Q is a measure of the distortion of the Cu square plane, for example for a rhombic distortion $Q = 2(d_1-d_2)/(d_1+d_2)$ where $d_1$ and $d_2$ are the Cu-O(planar) distances. The orthorhombic crystal structure in superconductors can be associated with the stabilized JT distortion of the Cu square plane at low temperature driven by g(Q). The variation of the dimpling angle $\beta$ or the Cu displacement from the oxygen plane in different Cu sites gives a modulation of the electron lattice interaction, via h($\beta$), and therefore it could indicate the dynamic coupling of the electronic states with lattice modes with wave-vector $k \neq 0$. The Cu-O(apical) distance modulates the electron lattice coupling $\lambda$ via the induced variation of the energy splitting $\Delta_{JT}$ which is maximum for infinite Cu-O(apical) distance and it decreases by shortening the Cu-O(apical) distance. From the experimental determination of the Cu site configurations in the cuprate superconductors we think that it is possible to formulate the hypothesis that the Cu site distortions indicate the formations of polarons involving the dimpling of the in plane oxygens and the variation of the Cu-O(apical) distance. The formation of polarons has been indicated by photoinduced infrared absorption results[23]

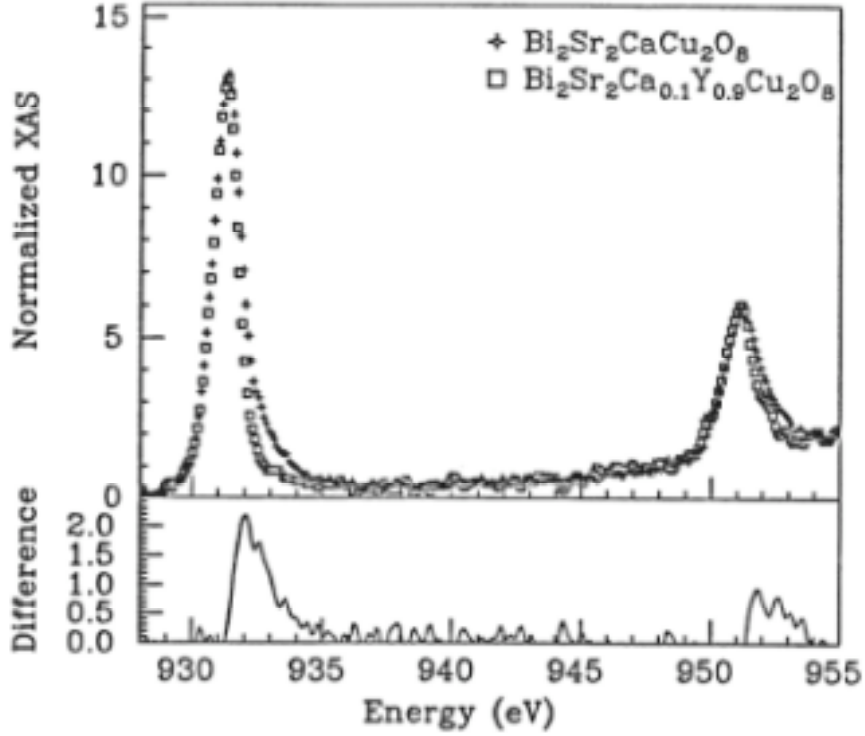

**FIG. 6.** The **E//ab** polarized Cu $L_3$ x-ray absorption spectra (XAS) of $Bi_2Sr_2CaCu_2O_8$ and that of the insulating $Bi_2Sr_2Ca_{0.1}Y_{0.9}Cu_2O_8$ crystal and their difference (lower panel).

The superstructure can be razionalized as evidence for ordering of polarons along one direction in the $CuO_2$ plane. The modulation of the Cu displacement and of the Cu-O(apical) distance giving the superstructure indicates different Cu sites where the electron lattice coupling is large (large β angle and short Cu-O(apical) distance) and other Cu sites where the electron lattice coupling is small (low β angle and long Cu-O(apical) distance). The electronic states in the $CuO_2$ plane appears therefore formed by two components: the first component is associated with the linear arrays of Cu sites with the long Cu-O(apical) distance along the 010 direction (see Fig.2) where the electron lattice coupling is weak and the second component with large electron lattice interaction that can be associated with the pseudo Jahn Teller polarons. The local density calculations show that a large electron phonon interaction is expected along the (110) direction, i.e. along the Cu-O-Cu lines, and it is expected to be enanched by the dimpling angle β.[24] The relevance for superconductivity of the vibration modes mixing the $3d_{x^2-y^2}$ with $3d_{3z^2-r^2}$ hole states is demonstrated by the Raman results[25]. The mode at 335 cm$^{-1}$ corresponding to out-of-phase vertical vibration (with $B_{1g}$ symmetry) of two oxygen atoms in the $CuO_2$ plane, that becomes soft at $T_c$ has the symmetry of the out of phase vibration ofte the O(P ions along the c axis shown in Fig. 5, mixing the $3d_{x^2-y^2}$ with $3d_{3z^2-r^2}$ states. The asymmetric Fano line shape of this Raman line is an indication of the interaction of this mode with electronic transitions.

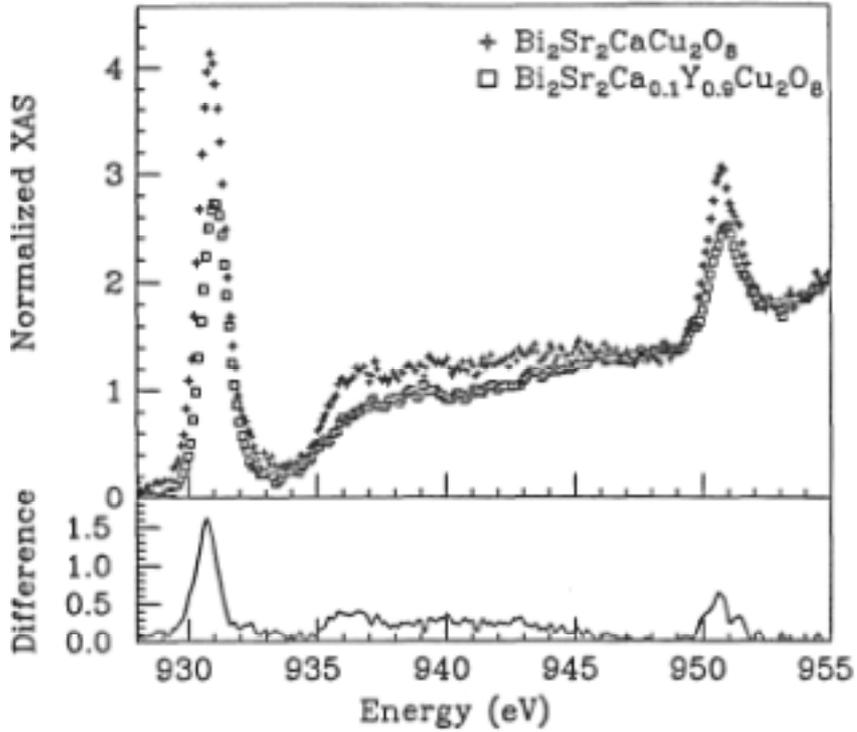

**FIG. 7.** The **E//c** polarized Cu $L_3$ x-ray absorption spectra (XAS) of $Bi_2Sr_2CaCu_2O_8$ and of the insulating $Bi_2Sr_2Ca_{0.1}Y_{0.9}Cu_2O_8$ crystal and their difference, lower panel.

### 3. The modulation of the electronic structure

The single hole per Cu site in the insulating parent system is frozen in the $\underline{3d_{x^2-y^2}}$ symmetry[26]. In high $T_c$ superconductors the doping introduces additional δ hole states per Cu ion in the insulating system resulting in 1+δ holes per Cu ion. The $3d^9\underline{L}$ character of additional hole states δ was found by x-ray absorption [27] and it is now well established. These states have a non negligible $a_1$ symmetry i.e. are given mixing the Cu $\underline{3d}_{3z^2-r^2}$ orbital with the combination of O(planar) $2p_{x,y}$ orbital with $a_1$ symmetry $\underline{L}(a_1)$, and O(apical) $2p_z$ orbital character[27-28] beyond the majority of carriers having $\underline{3d}_{x^2-y^2}$ and $\underline{L}(b_1)$ character.

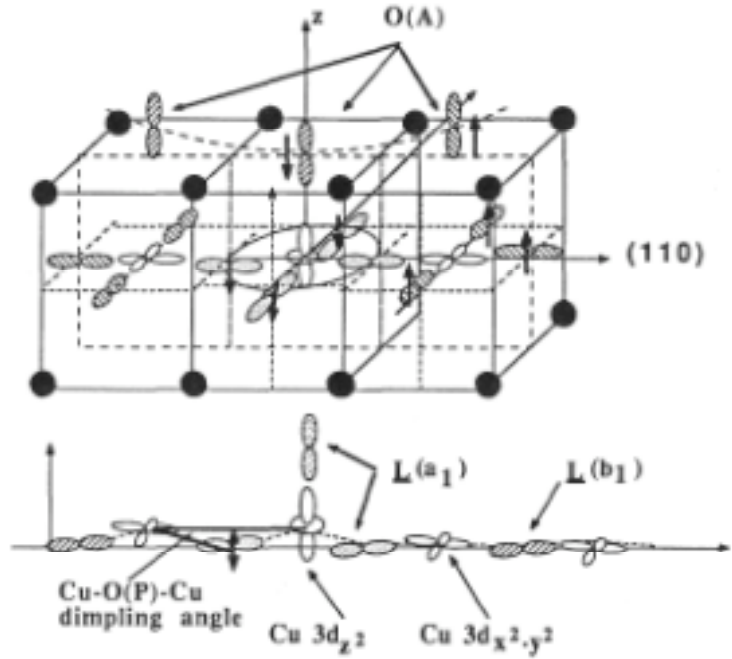

**FIG. 8** Pictorial view of the correlation between the modulation of the Cu site structure configurations and the modulation of the symmetry of the Cu 3d holes and of the oxygen 2p molecular orbitals with $a_1$ and $b_1$ symmetry.

The presence of some Cu 3d holes with $a_1$ symmetry in metallic systems was associated with the formation of pseudo Jahn Teller (JT) polarons characterized by short Cu-O(apical) distance and distortions of the CuO$_4$ square plane [28] and their relevance for some pairing mechanisms was discussed by Askenazi [29].

Theoretical calculations [30, 31] show that the energy separation between the $b_1$ and $a_1$ states is modulated by the Cu-O(apical) distance therefore at the sites with short Cu-O(apical) bond the Cu 3d $m_\ell = 0$ component of the electronic carriers, or with $a_1$ molecular symmetry, is expected to increase. The stability of a Jhan Teller (called also anti-JT) polaron associated with $a_1$ states in LaSrCuO system was discussed recently [32,24]. We have measured the variation of the polarized Cu L$_3$ XAS absorption line of a Bi 2:2:1:2 superconducting single crystal compared with the Yttrium doped insulating system as it was previously reported for the LaSrCuO[33] superconductor and for a single layer Bi 2:2:0:1 system[34].

The Cu $L_3$ XAS spectra of the insulating system probes the single hole per Cu ion therefore the difference spectra probe the $3d^9\underline{L}$ states. In the Bi 2:2:1:2 single crystal we have found a total of $\delta = 0.19\pm 0.2$ $d^9\underline{L}$ states from the difference of the unpolarized spectra. The difference between the polarized **E//ab** spectra of the metal and of the insulating system, reported in Fig. 6, shows that $0.15 \pm .02$ added holes have the $3d_{x^2-y^2}\underline{L}(b_1)$ character. In Fig. 7 the difference spectrum between the **E//c** XAS spectra of the metal and the insulating system show the presence of the other $0.04 \pm .02$ states with $\underline{3d}_{3z^2-r^2}\underline{L}(a_1)$ character.

The final states due to the transition Cu $2p^6\underline{3d}_{3z^2-r^2}\underline{L}(a_1)$ -> Cu $2p^53d^{10}\underline{L}(a_1)$ in **E//c** difference XAS spectra are at lower energy (at about 931.1 eV in the lower panel of Fig. 7) than the final states due to the Cu $2p^6\underline{3d}_{x^2-y^2}\underline{L}(b_1)$ -> Cu $2p^53d^{10}\underline{L}(b_1)$ that appear in the **E//ab** XAS difference spectra, at about 932 eV in the lower panel of Fig. 6. Because the absorption line due to the Cu 2p -> 3d transition is a bound excitonic state, the higher energy position of the $2p^53d^{10}\underline{L}(b_1)$ final states indicates mobile carriers with $b_1$ symmetry screening the core hole. On the contrary the fact that the line due to the Cu $2p^53d^{10}$ $\underline{L}(a_1)$ final state is at the same energy as that of the absorption line in the insulating system indicates that the states with $a_1$ symmetry are more localized.

Therefore it is possible to associate the more delocalized holes with $b_1$ symmetry with the linear arrays along the b axis in Bi2212 with the Cu site configurations with long Cu-O(apical) distances. The more localized states, induced by doping, with partially $a_1$ symmetry can be associated with the electronic states with the modulated Cu site configurations along the Cu-O-Cu-O direction with the short and long Cu-O(apical) distances A pictorial view of the proposed correlation between the variation of the local symmetry of the electronic carriers and the variation of the Cu site structure is shown in Fig. 8.

## 4. Conclusions

The present work supports the formation of pseudo Jahn Teller polarons driven by the electronic states, induced by doping, with partially $a_1$ symmetry characterized by Cu sites with the short Cu-O(apical) distance and Cu displacement from the in plane oxygens. These Cu sites are distributed in the $CuO_2$ plane with a regular arrangement of linear arrays in the (010) direction separated by linear arrays of sites with long Cu-O(apical) distance.

This new scenario suggests pairing mechanisms for high $T_c$ superconductivity in the presence of two components of the electronic structure: the more delocalized component formed by $\underline{3d}_{x^2-y^2}\underline{L}(b_1)$ states and the more localized states with partially $a_1$ symmetry. A possible pairing mechanism in this scenario see the pairing of the delocalized carriers with $b_1$ symmetry mediated by excitations in the more localized $a_1$ component. Therefore the system can be described as formed by an infinite number of parallel interacting chains of itinerant states separated by linear chains of localized states.

It is interesting to remark that the wave length of the superstructure $\lambda$ is close to the coherence length $\xi_{ab} = 26\pm5$ Å (values ranging from 21 Å to 31 Å are reported in the literature[35]) in Bi 2:2:1:2. Because $1/\xi_{ab}$ gives the width of the momentum distribution involved in the pairing it is possible that the excitation exchanged in the pairing has the wave-vector close to $1/\lambda$.

The regular array of stripes of polarons, separated by $\lambda$, seems to be correlated with the wave-vector of the carriers at the Fermi level $k_F \sim 2\pi/\lambda_F$. More extensive investigations on the relation between the superstructure and the shape of the Fermi surface as function of doping will clarify if the structural instability giving the superstructure is driven by electronic structure.

The $k_F \cdot \xi_{ab} \sim 2\pi$ relation that is valid for high $T_c$ superconductors seems to be related with the ordering in the real space with the spacing $\lambda$ between the stripes of polarons.

An interesting aspect of the hypothesis that the resulting excitation exchanged by the carriers is related to the electron-lattice interaction of the pseudo Jahn Teller type is that it will have both phononic, excitonic and magnetic characters.

I would like to thank S. Della Longa, A.M. Flank, P. Lagarde, I. Pettiti, M. Pompa, P. Porta S. Turtù, D. Udron, and A. Di Cicco for experimental help and G. Calestani, C. Castellani, F. De Martini, C. Di Castro, M. Grilli, and L. Pietronero for useful discussions.

**APPENDIX Recorded discussion at the first workshop on Phase Separation in Cuprate Superconductors held in Erice 1992 and published in the book edited by K. A. Müller and G. Benedek World Scientific., Singapore, 1992**

**Antonio Bianconi.** At this meeting finally the disagreement between two classes of structural probes on the Cu local structure was solved for two classes of high Tc superconductors. A long standing controversy has been going on for about five years between X-ray absorption spectroscopy (EXAFS and XANES), pulsed neutron diffraction, experimental methods that probe the local structure, and crystallography investigations (neutron and x-ray diffraction), which probe the long range order, on the presence of two positions for the apical oxygen O(4) in $YBa_2Cu_3O_{7-\delta}$ and second, on the anomalies of the local structure at the normal to superconductor phase transition, which have been found by XANES, EXAFS and pulsed neutron scattering but they have not been found by crystallography.

At this meeting we have presented the evidence that the Cu sites in $Bi_2Sr_2CaCu_2O_{8+\delta}$ are not homogeneous. We can distinguish two main types of Cu sites:

**Type 1)**, where the Cu ion is displaced by about 0.2 Å above the distorted square plane of the planar oxygen's, and the Cu-O(apical) distance is shorter ~ 2.3 Å and

**Type 2)**, where the Cu ion is close to the plane of planar oxygen's and the Cu-O(apical) distance is longer, about ~ 2.5 Å.

It is relevant that the two types of Cu site configurations are similar in $YBa_2Cu_3O_{6.5}$ and $Bi_2Sr_2CaCu_2O_{8+\delta}$.

$Bi_2Sr_2CaCu_2O_{8+\delta}$ provides the ideal system to compare the results of experimental investigations probing the local order or the long-range order. In fact this system exhibits a well-defined 4.75 x 1 superstructure and therefore the structure of the different Cu sites can be obtained also by diffraction methods if they solve the superstructure. The EXAFS investigation of the local structure by our group has shown the presence of two Cu-O(apical) distances and of two different Cu sites. These results are in agreement with the diffraction works that solved the superstructure, but cannot be compared with the other works that report the average structure. The superstructure of $YBa_2Cu_3O_{6.5}$ solved by Burlet et al. shows the presence of two different Cu-O(apical) distances, in agreement with EXAFS data of Conradson et al.. Therefore at this meeting two cases, $YBa_2Cu_3O_{6.5}$ and $Bi_2Sr_2CaCu_2O_{8+\delta}$, have been presented where the experimental methods probing the long and short order are in agreement.

I think that also the controversy on the Cu-O(apical) distance in $YBa_2Cu_3O_{7-\delta}$ can be interpreted as due to superstructure in a small domains or with the dynamic character of the superstructure that cannot be seen by diffraction. The EXAFS experiments probe both the static and dynamic disorder because of the very shot measuring time, $10^{-16}$ sec, therefore the anomalies at the superconducting phase transitions can be assigned to the different dynamics of the two types of Cu sites. Egami et al. by pulsed neutron diffraction have observed the increasing probability for Cu sites of type 2 below the critical temperature. I think that we have to look at the superstructure as having both static and dynamic character.

How common is the presence of a superstructure with linear arrays of Cu sites of types 1 and 2 in the high Tc superconductors? At this meeting Grenier et al. have reported the presence of a superstructure in $La_2CuO_{4+x}$ (0.6<x<0.8) similar to that of $Bi_2Sr_2CaCu_2O_{8+\delta}$. On the other side no superstructure has been solved up to now in several high $T_c$ superconductors like in $La_{2-x}Sr_xCuO_4$ and in $YBa_2Cu_3O_7$ however it is possible that the size of the ordered domains can be as small as the superconducting coherence length ~20 Å that they escape to be detected by diffraction methods.

I would like to stress that the aspect that seem to me more relevant for the physics of high $T_c$ superconductivity is the fact the alternate lines of Cu sites are formed in the $CuO_2$ plane. The new scenario that appear now for the $CuO_2$ plane is similar to the shape of a corrugated iron foil. Linear arrays of type 1 Cu sites configurations and of type 2 Cu sites are present in the plane. Therefore the plane is no more isotropic and the carriers move more freely in one direction than in the other direction.

**Carlo Di Castro**. I would like to make a general comment. First of all, the involvement of the apex oxygen changes the symmetry of the holes in the plane. This means that there is a change from the simplified picture we had for many years from the Zhang and Rice singlet, in the sense that there is a relevant part of holes with $a_1$ symmetry. Concerning the number of bands there is consensus that only one band crosses the Fermi energy. However this band is made by holes with different local symmetry. This means that they respond differently to local distortions and to the effective potential which comes out from different techniques and mechanisms. A have a question for experimentalists. As Bianconi said the superstructure where the apex oxygen is involved has a complicated behavior where several crystallographic parameters are modulated involving also tilting of the square pyramids and so on. Now you have to be good with theorists. We cannot introduce in the theory many types of distortions. To get information on the holes on the plane one has to identify the parameter modulated by the superstructure that is relevant for local distortions and symmetry. This may be helpful for the selection of the effective pairing.

**Antonio Bianconi**. With the present experimental information, the main aspects of the modulation of the Cu site configurations that I think are relevant for the pairing mechanisms are the following.

1) The Cu sites can be identified by the long or short Cu-O(apical) distance. This modulation induces that the local symmetry $a_1$ of the holes in the plane is larger for type 1 sites and smaller for type 2 sites.

2) The different Cu sites form linear arrays. I think that the superstructure can be different in different systems, but the formation of linear arrays seem to be a common aspect of an instability driven by the electron lattice interaction in the planes.

3) The third point is that the period of the modulation $\lambda$ satisfies the condition $k_F \cdot \lambda \sim 2\pi$, where $k_F$ is the wave vector of the itinerant holes at the Fermi level in the direction of higher dispersion. Moreover it has been found that the superconducting coherence length $\xi$ is close to the modulation period. Therefore the condition $k_F \cdot \xi \sim 2\pi$ is satisfied.

I would like to suggest a pairing mechanism that is suggested by these new data. We have the presence of two components in the carrier spectrum in the planes. The *first* component is formed by *charges that moves freely mainly in one direction like the water running on the grooves of a corrugated iron foil*. The first type of carriers, with mostly $b_1$ symmetry and associated with Cu sites of type 2, move freely in one direction. The *second* type of charge carriers, with larger components of states with $a_1$ local symmetry, associated with the Cu sites of type 1, that have a strong electron lattice interactions and form probably *linear arrays of condensed polarons* separated by the modulation period $\lambda$. The pairing of the charges in the first set of states is mediated by virtual excitations with a wave vector $2\pi/\lambda$.